\documentclass[prx, twocolumn, superscriptaddress]{revtex4-2}
\usepackage{amsmath,amssymb,amsfonts}
\usepackage{algorithmic}
\usepackage{graphicx}
\usepackage{textcomp}
\usepackage{comment}
\usepackage{here}
\usepackage{braket}
\usepackage{physics}
\usepackage{wrapfig}
\usepackage{mathtools}
\usepackage{caption}
\newcommand{\cut}{\mathrm{cut}}

\def\BibTeX{{\rm B\kern-.05em{\sc i\kern-.025em b}\kern-.08em
    T\kern-.1667em\lower.7ex\hbox{E}\kern-.125emX}}
\begin{document}

\title{Noise Robustness of Quantum Relaxation for Combinatorial Optimization}

\author{Kentaro Tamura}
\email[]{cicero@keio.jp}
\affiliation{Department of Applied Physics and Physico-Informatics, Keio University, 3-14-1 Hiyoshi, Kohoku-ku, Yokohama, Kanagawa, 223-8522, Japan}

\author{Yohichi Suzuki}
\affiliation{Quantum Computing Center, Keio University, 3-14-1 Hiyoshi, Kohoku-ku, Yokohama, Kanagawa, 223-8522, Japan}

\author{Rudy Raymond}
\affiliation{IBM Quantum, IBM Japan, 19-21 Nihonbashi Hakozaki-cho, Chuo-ku, Tokyo 103-8510, Japan}
\affiliation{Quantum Computing Center, Keio University, 3-14-1 Hiyoshi, Kohoku-ku, Yokohama, Kanagawa, 223-8522, Japan}
\affiliation{Department of Computer Science, The University of Tokyo, 7-3-1, Hongo, Bunkyo-ku, Tokyo 113-0033, Japan}

\author{Hiroshi C. Watanabe}
\affiliation{Department of Chemistry, Graduate School of Science, Kyushu University, 744 Motooka, Nishi-ku, Fukuoka, 819-0395, Japan}
\affiliation{Quantum Computing Center, Keio University, 3-14-1 Hiyoshi, Kohoku-ku, Yokohama, Kanagawa, 223-8522, Japan}

\author{Yuki Sato}
\affiliation{Toyota Central R\&D Labs., Inc., 41-1, Yokomichi, Nagakute, Aichi, 480-1192, Japan.}
\affiliation{Quantum Computing Center, Keio University, 3-14-1 Hiyoshi, Kohoku-ku, Yokohama, Kanagawa, 223-8522, Japan}

\author{Ruho Kondo}
\affiliation{Toyota Central R\&D Labs., Inc., 41-1, Yokomichi, Nagakute, Aichi, 480-1192, Japan.}
\affiliation{Quantum Computing Center, Keio University, 3-14-1 Hiyoshi, Kohoku-ku, Yokohama, Kanagawa, 223-8522, Japan}

\author{Michihiko Sugawara}
\affiliation{Quantum Computing Center, Keio University, 3-14-1 Hiyoshi, Kohoku-ku, Yokohama, Kanagawa, 223-8522, Japan}

\author{Naoki Yamamoto}
\affiliation{Department of Applied Physics and Physico-Informatics, Keio University, 3-14-1 Hiyoshi, Kohoku-ku, Yokohama, Kanagawa, 223-8522, Japan}
\affiliation{Quantum Computing Center, Keio University, 3-14-1 Hiyoshi, Kohoku-ku, Yokohama, Kanagawa, 223-8522, Japan}

\begin{abstract}
QRAO (Quantum Random Access Optimization) is a relaxation algorithm that reduces the number of qubits required to solve a problem by encoding multiple variables per qubit using QRAC (Quantum Random Access Code). Reducing the number of qubits is a common way of dealing with the impact of noise on a quantum algorithm. Our interest lies in the impact of noise on the quality of the binary solution of QRAO, which is unknown. We demonstrate that the mean approximation ratio of the (3, 1)-QRAC Hamiltonian, i.e., the Hamiltonian utilizing the encoding of 3 bits into 1 qubit by QRAC, is less affected by noise compared to the Ising Hamiltonian used in quantum annealer and QAOA (Quantum Approximate Optimization Algorithm). Based on this observation, we discuss a plausible mechanism behind the robustness of QRAO under depolarizing noise. Finally, we assess the number of shots required to estimate the values of binary variables correctly under depolarizing noise and show that the (3, 1)-QRAC Hamiltonian requires less shots to achieve the same accuracy compared to the Ising Hamiltonian.
\end{abstract}
\maketitle

\section{Introduction}
\label{sec:introduction}
Combinatorial optimization is the task of finding an optimum value of a function defined on some typically finite domain~\cite{grotschel1995combinatorial}. The task has a wide range of applications ranging from industry~\cite{azad2022solving, kurowski2023application} to finance~\cite{Egger2020, wang2020}. Quantum algorithms for optimization such as VQE (Variational Quantum Eigensolver)~\cite{vqe} and QAOA (Quantum Approximate Optimization Algorithm)~\cite{qaoa, qaoa2} share a common concern when executed on a noisy intermediate-scale quantum (NISQ) device, which is the problem of scalability~\cite{abbas2023quantum}. On NISQ devices, the number of sequential gate operations while sustaining a coherent quantum state is restricted by noisy operations and the limited number of qubits~\cite{moll2018quantum}. One way to deal with scalability is to reduce the number of qubits in a circuit, for example by cutting a large circuit into smaller sub-circuits with fewer qubits and less sequential gate operations~\cite{tang2021cutqc}. In quantum algorithms for optimization, the number of qubits employed is determined by the encoding of a problem, which is the mapping of classical variables onto qubits. Various encodings have been proposed to achieve a more efficient encoding~\cite{Tan2021qubitefficient, yano}. Among them is the QRAO (Quantum Random Access Optimization)~\cite{qrao} algorithm proposed by Fuller et al., which utilizes QRAC (Quantum Random Access Code)~\cite{ambainis2002dense} to encode multiple binary variables per qubit, thereby reducing the number of qubits required for problem mapping.

QRAO\cite{qrao, qrao_entanglement, qrao_tetrahedron, 2024arXiv240302045K} differs from algorithms such as QAOA in that it involves a process called quantum state rounding. Quantum state rounding is the mapping of the candidate state obtained by methods such as VQE onto a binary solution. Because a candidate state perturbed by noise may still be mapped to the same binary solution, we have the intuition that the solution of QRAO may be robust to noise. In this paper, we are interested in the effect of noise on the quality of the binary solution obtained by QRAO. We encode the same combinatorial optimization problems onto two Hamiltonian: the QRAC Hamiltonian and the Ising Hamiltonian, use the same ansatz and optimizer for VQE to obtain the candidate states, and compare the binary solutions characterized by the approximation ratio. The approximation ratio is the ratio between the cost function value of the binary solution at hand and the optimal binary solution.

QRAC was first proposed in the context of communication in order to encode as many classical bits per qubit. The central idea was to exceed the Holevo bound~\cite{holevo1973} that forbids encoding $m$ bits into less number of qubits without information loss by allowing a possibility of decoding the wrong bit. The encoding of $m$ binary variables on $n$ qubits with decoding probability $p$ is denoted as $(m,n,p)$-QRAC~\cite{ambainis2002dense}.
There are $(2,1,0.85)$-QRAC, $(3,1,0.78)$-QRAC~\cite{ambainis2002dense,hayashi20064}, and several other constructions of $(m,2,p>1/2)$-QRACs~\cite{Liab_tr__2017,imamichi2018constructions,manvcinska2022geometry}. For simplicity, the probability $p>1/2$ in $(m,n,p)$-QRAC is omitted and written as $(m,n)$-QRAC from now on. While the effect of noise on QRAC in its original context has been reported~\cite{marques2022quantum}, the effect of noise on QRAC in the context of optimization is unknown. 

In the present study, we evaluate the effect of noise on the quality of the binary solution of QRAO with respect to the approximation ratio. We demonstrate through simulation using a noiseless device that as the problem-size increases, the mean approximation ratio resulting from the (3, 1)-QRAC Hamiltonian exceeds the mean approximation ratio resulting from the Ising Hamiltonian. The simulation results under fake noise~\cite{qiskit} show that the mean approximation ratio obtained by the (3, 1)-QRAC Hamiltonian is more robust to noise than in the Ising Hamiltonian case. We provide a theoretical explanation for the effect of noise on the mean approximation ratio of QRAO by assuming depolarizing noise. Finally, we derive the order of shots required in Pauli rounding to achieve a given successful decoding probability under depolarizing noise.

\section{Preliminary}
\subsection{Maximum cut (MaxCut) problem}
\begin{figure}[t]
    \includegraphics[width=88mm]{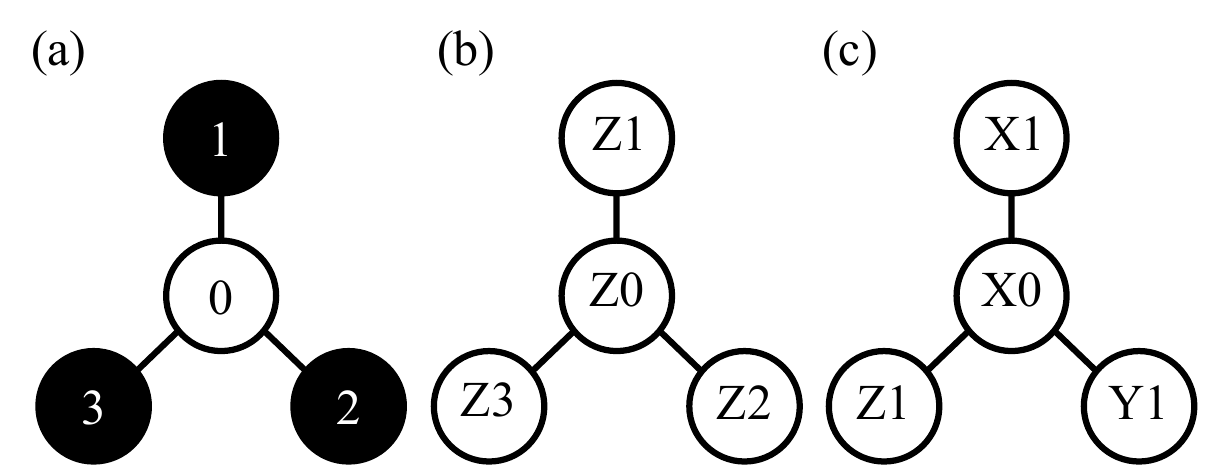}
    \caption{(a) One of the optimal configurations of the MaxCut problem for a 4-node graph. (b) An example of a Ising encoding for a 4-node graph using 4 qubits, where $Zi$ corresponds to the Pauli matrix $Z$ on qubit index $i$. (c) An example of a (3, 1)-QRAC encoding for a 4-node graph using 2 qubits, where $Pi$ corresponds to the Pauli matrix $P\in\{X,Y,Z\}$ on qubit number $i$.\label{fig:petersen}}
\end{figure}

In this paper, we deal with the unweighted MaxCut problem. The MaxCut problem is an NP-hard combinatorial problem involving undirected graphs~\cite{Karp1972}. Given an undirected graph $G$ with $|V|$ nodes labeled $v_i$ and $|E|$ edges labeled $e_{i,j}$, the objective of the MaxCut problem is to find a configuration $m_i\in \{0, 1\}$ that maximizes the cost function
\begin{align}
\max_{m\in\{0, 1\}^{|V|}}\cut(m),
\label{eq:maxcut}
\end{align}
where
\begin{align*}
\cut(m)\coloneqq \frac{1}{2}\sum_{e_{i,j}\in E} (1 - (-1)^{m_i + m_j}).
\end{align*}
For example, one of the optimal solutions to the MaxCut problem for a 4-node graph is shown in Fig.~\ref{fig:petersen}~(a), where 3 out of 3 edges are included in the cut.
\noindent The solution accuracy is evaluated by the approximation ratio defined by the ratio between the obtained cut value ($\cut(m)$) and the optimal cut value ($\cut(m^*)$). 
The approximation ratio represented by $\gamma$ is a real number ranging from 0 to 1. 
For example, the approximation ratio of the output in Fig.~\ref{fig:petersen}~(a) is $\gamma = 3/3 = 1.0$. 
\subsection{Quantum Random Access Optimization (QRAO)}
QRAO~\cite{qrao} is  a relaxation-based optimization algorithm that uses QRAC to solve a binary optimization problem.
The use of QRAC enables us to save the number of qubits to one-third as many qubits as the number of binary variables (bits). Decoding the binary solution from a qubit requires a specific measurement procedure rather than a simple measurement in the computational basis.
QRAO consists of three steps: encoding, optimization and rounding.
In encoding, we construct the QRAC Hamiltonian, which encodes the binary optimization problem in a relaxed manner.
In optimization, VQE is carried out based on the QRAC Hamiltonian.
The binary solution is then estimated from the resulting quantum state through a measurement process termed quantum state rounding. In this section, we overview each step of the algorithm with the MaxCut problem as an example.

\subsubsection{Encoding}
In this paper, we define {\it encoding} as the embedding of classical bits into qubits. In the conventional Ising-type formulation~\cite{qaoa}, the classical bit 0 is encoded to $\ket{0}$ and 1 to $\ket{1}$, which can be viewed as the $i$-th node of the graph is assigned to the Pauli matrix $Z_i$ supported by the $i$th qubit.
Hence, the score of $e_{i,j}$ is defined as
$\frac{1}{2}(I - Z_iZ_j)$.
As a result, the MaxCut problem of a graph $G$ is equivalent to the maximization of the mean value of the following Hamiltonian:
\begin{align}
H = \frac{1}{2}\sum_{e_{i, j}\in E} (I - Z_iZ_j).
\label{eq:qubo_hamiltonian}
\end{align}
\begin{figure}[t]
    \centering
    \includegraphics[width=88mm]{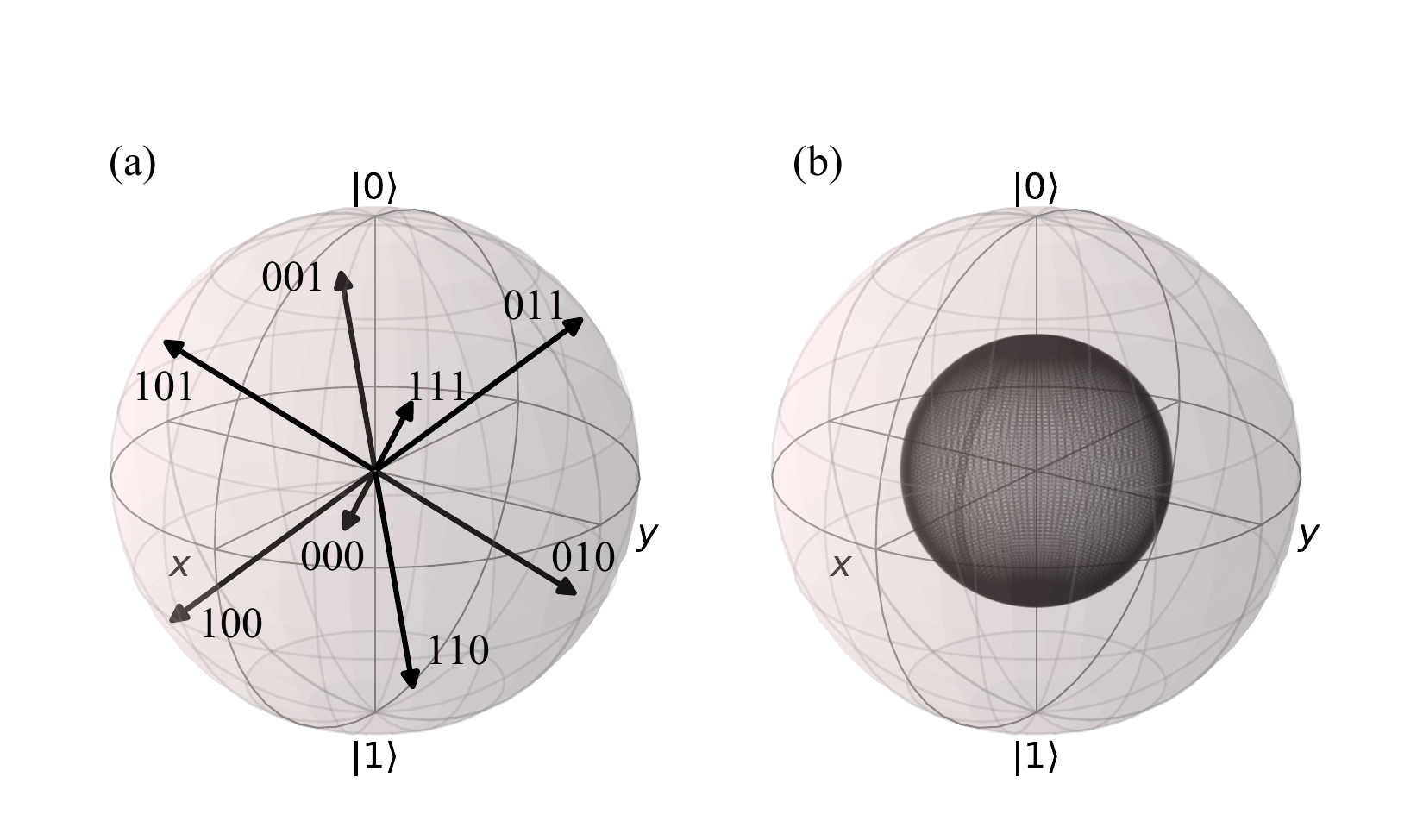}
    \caption{(a) Bloch sphere representation of (3,1)-QRAC. (b) Bloch sphere representation of the depolarizing channel with error probability 0.5.}
    \label{fig:bloch}
\end{figure}
In the QRAC formulation, the classical bits $(x_1, x_2, x_3)$ are encoded as
\begin{align}
&f(x_1, x_2, x_3) =\nonumber \\
&\frac{1}{2}\left(I + \frac{1}{\sqrt{3}}((-1)^{x_1}X + (-1)^{x_2}Y + (-1)^{x_3}Z)\right),
\end{align}
where $X, Y$, and $Z$ are Pauli matrices, $I$ the identity matrix, and $x_1, x_2, x_3\in \{0, 1\}$. 
The encoded states are plotted at the vertices of a cube in the Bloch sphere as in Fig.~\ref{fig:bloch}~(a). 
One can assign at most three nodes to each qubit with the constraint that adjacent nodes must be assigned to different qubits. The Hamiltonian is constructed as
\begin{align}
H = \frac{1}{2}\sum_{e_{i, j}\in E} (I - 3P_{i}P_{j}),
\label{eq:qrac_hamiltonian}
\end{align}
where $P_{i}$ corresponds to the Pauli matrix assigned to the $i$-th node. For a candidate state $F(m)$ which is a product state of $f$, we have~\cite{qrao}
\begin{align*}
\Tr(F(m)H) = \cut(m).
\end{align*}
Note that as $H$ is a relaxed Hamiltonian, the expectation value may exceed the maximum cut value. We expect that maximizing the expectation value of the candidate state with respect to the Hamiltonian results in a closer state to $F(m)$.
\subsubsection{Optimization}
\begin{figure}[t]
    \centering
    \hspace{-10mm}
    \includegraphics[width=88mm]{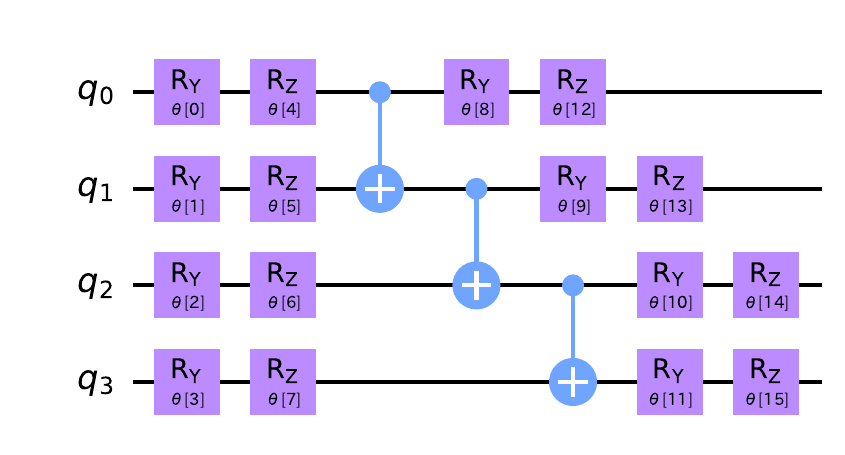}
    \caption{A 4-qubit single-layer hardware-efficient ansatz with linear entanglement.}
    \label{fig:ansatz}
\end{figure}

In the optimization step, the expectation value of the QRAC Hamiltonian is maximized by varying the quantum state via variational methods such as VQE or QAOA. 
In the present study, we carry out VQE with the hardware-efficient ansatz to obtain the candidate state. 
An example of the ansatz with 4 qubits is shown in Fig.~\ref{fig:ansatz}.
Although we are aware that Free axis selection\cite{watanabe2023optimizing} or Free quaternion selection\cite{wada2022simulating,wada2022full} methods are generally upper compatible optimizers with the NFT algorithm (also termed Rotosolve) ~\cite{nft, ostaszewski2021structure}, they do not necessarily bring benefits in the Ising Hamiltonian, and thus we consistently used the NFT (Nakanishi-Fujii-Todo) algorithm for the Ising and QRAC Hamiltonian.

\subsubsection{Quantum State Rounding}
The maximum-eigenvalue eigenstates of the Ising Hamiltonian directly correspond to the classical solution because the Ising Hamiltonian is diagonal in the computational basis. 
To obtain the classical solution, therefore, one needs only to apply measurement in the computational basis $\{\ket{0}, \ket{1}\}$. 
In contrast, the eigenstates of the (3, 1)-QRAC Hamiltonian are not necessarily diagonal in the computational basis; the obtained eigenstates are in entanglement and superposition.
To obtain the classical solution, a procedure called quantum state rounding is required. 
QRAO has two methods of rounding: Pauli rounding and magic state rounding\cite{qrao}. 

\paragraph{Pauli rounding}
In Pauli rounding, the classical bits are decoded by performing expectation value estimation with Pauli matrices as observables. Given a candidate state $\rho$, the classical bit is decoded by estimating the sign of the trace value $\Tr(P(v_i)\rho)$, where $P(v_i)$ denotes the Pauli matrix corresponding to the node $v_i$. If the trace value is positive, $+1$ is assigned, if negative, $-1$ is assigned, and if 0, a bit of 0 or 1 is assigned uniformly at random.  
For example, when the corresponding qubit index and the Pauli string for node $v_0$ are 0 and $X$, respectively, $\Tr[(X \otimes I) \rho]$ is used to decode the classical bit of $v_0$ (see Fig.~\ref{fig:petersen}(c)). Here, the Pauli matrix on the far left corresponds to the qubit with the smallest index.
\paragraph{Magic state rounding}
In magic state rounding, each qubit of the candidate state is measured along the following four bases uniformly at random:
\begin{align*}
\mu_1^\pm &= \frac{1}{2}
\left(
I \pm \frac{1}{\sqrt{3}}(X + Y + Z)
\right),\\
\mu_2^\pm &= \frac{1}{2}
\left(
I \pm \frac{1}{\sqrt{3}}(X - Y - Z)
\right),\\
\mu_3^\pm &= \frac{1}{2}
\left(
I \pm \frac{1}{\sqrt{3}}(-X + Y - Z)
\right),\\
\mu_4^\pm &= \frac{1}{2}
\left(
I \pm \frac{1}{\sqrt{3}}(-X - Y + Z)
\right).
\end{align*}
This procedure is equivalent to mapping each qubit to one of the 8 states shown in Fig.~\ref{fig:bloch}~(a). Once the qubits have been assigned one of 4 bases, all qubits are measured once along the assigned basis. This is repeated for a number of times, and the solution with the highest approximation ratio becomes the final output. The lower bound for the expected approximation ratio for magic state rounding is known to be~\cite{qrao}
\begin{align}
    \mathbb{E}(\gamma) 
    &= \mathbb{E}\left[\frac{\cut(m)}{\cut(m^*)}\right]\nonumber\\
    &= \frac{\mathbb{E}\left[\Tr(\mathcal{M}^{\otimes n}(\rho)H)\right]}{\Tr( F(m^*)H)}\geq \frac{5}{9},
\end{align}
where $F(m^*)$ is a map that achieves $\Tr(F(m^*)H) = \cut(m^*)$, and $m^*$ the optimal configuration. Here, $\mathcal{M}^{\otimes n}(\rho)$ is defined as the $n$-qubit state resulting from a single shot of magic state rounding, which is a product state of the eight QRAC states shown in Fig.~\ref{fig:bloch}~(a) in the case of (3, 1)-QRAC~\cite{qrao}.

\section{Results and Discussion}
Our main result is shown in Fig.~\ref{fig:qrac_simulation}. We will henceforth refer to the (3, 1)-QRAC Hamiltonian as the QRAC Hamiltonian. The approximation ratio corresponding to the QRAC Hamiltonian is obtained by Pauli rounding unless specified otherwise.
\subsection{Simulation results of QRAO under noise}
To examine the effect of noise on the approximation ratio of QRAO, we solved the MaxCut problem for random 3-regular graphs with the QRAC Hamiltonian and the Ising Hamiltonian using candidate states obtained by using the following devices:
\begin{itemize}
\item Statevector simulator without noise,
\item Statevector simulator with fake noise,
\item Statevector simulator with depolarizing noise with error probability $1~\%$ on the controlled-NOT gate.
\end{itemize}

\begin{figure*}[t]
    \centering
    \hspace{-8mm}
    \includegraphics[width=181mm]{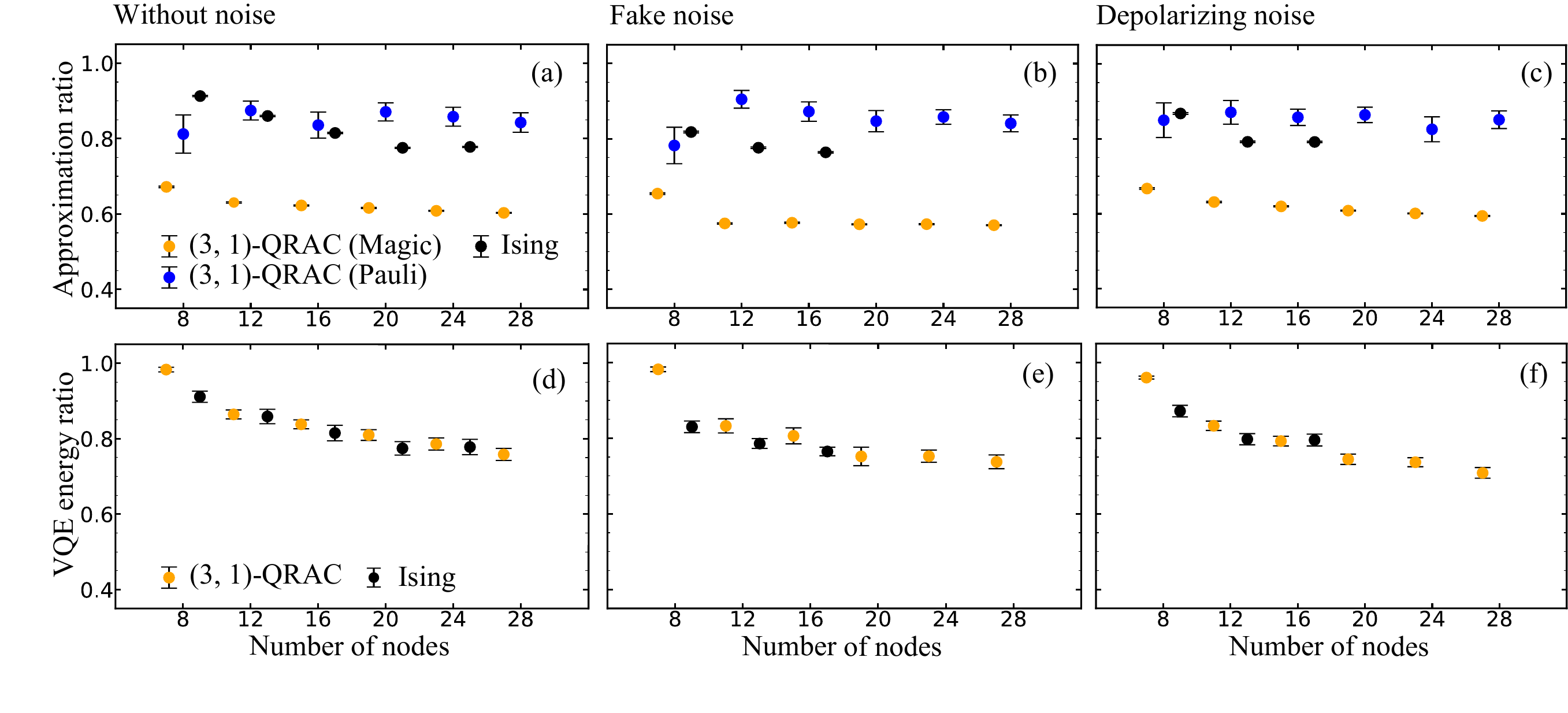}
    \caption{
    (a) Rounding results of candidate states obtained via VQE executed with 1024 shots for each Pauli terms to estimate the energy expectation value under no noise. (b) Rounding results of candidate states obtained via VQE executed with 1024 shots for each Pauli term and with the fake backend FakeMumbaiV2~\cite{qiskit} provided by IBM. (c) Rounding results of candidate states obtained via VQE executed with 1024 shots for each Pauli term and with noisy controlled-NOT gates under the influence of depolarizing noise with error probability $1~\%$. (d) The mean of the VQE energy ratio corresponding to the candidate states of (a). (e) The mean of the VQE energy ratio corresponding to the candidate states of (b). (f) The mean of the VQE energy ratio corresponding to the candidate states of (c)}
    \label{fig:qrac_simulation}
\end{figure*}

\noindent Fake noise refers to a noise model that mimics the behavior of a real device by combining the single qubit depolarizing error, single qubit thermal relaxation error, two-qubit depolarizing error, and the single qubit readout error, whose parameters are tuned based on real system snapshots~\cite{qiskit}. For both types of Hamiltonian, the candidate states were prepared via VQE with 3 layers of the hardware efficient ansatz shown in Fig.~\ref{fig:ansatz}~(b) and 2 parameter sweeps with the NFT algorithm. We employed the linear entanglement ansatz, in which case the controlled-NOT gate depth is $n-1$ for $n$ qubits. Noisy simulations for graphs with 20 nodes or more via the Ising Hamiltonian could not be executed in a reasonable amount of time. The mean approximation ratios obtained by the respective devices are shown in Fig.~\ref{fig:qrac_simulation}~(a), (b), and (c), where the error bars represent the 95~\% confidence intervals. The ratio between the achieved energy of the candidate state and the maximum eigenvalue of the Hamiltonian is shown in Fig.~\ref{fig:qrac_simulation}~(d), (e), and (f), along with the 95~\% confidence intervals. For each number of nodes, 50 random 3-regular graphs were solved.

Fig.~\ref{fig:qrac_simulation}~(a) is the simulation results of the statevector simulator without noise. It shows that as the problem-size characterized by the number of nodes increases, the mean approximation ratio of the QRAC Hamiltonian exceeds that of the Ising Hamiltonian. The mean approximation ratio of the Ising Hamiltonian shows a significant decline under the level of significance of 5\% from 8 to 20 nodes, while the mean approximation ratio of the QRAC Hamiltonian does not. From Fig.~\ref{fig:qrac_simulation}~(d), we observe that there is no significant difference in VQE energy ratio between the candidate states of the QRAC Hamiltonian and the Ising Hamiltonian at 20 nodes or more. However, the mean approximation of the QRAC Hamiltonian at 20 nodes is significantly higher than that of the Ising Hamiltonian. This indicates that the QRAC tends to yield a higher approximation ratio with respect to the VQE energy ratio.

Fig.~\ref{fig:qrac_simulation}~(b) shows the mean approximation ratios obtained under fake noise. By comparing Fig.~\ref{fig:qrac_simulation}(a) and (b), we find that the mean approximation ratio of the Ising Hamiltonian shows a significant decline when subjected to noise, while the mean approximation ratio of the QRAC Hamiltonian does not. This implies that the mean approximation ratio of the QRAC Hamiltonian via Pauli rounding is robust to noise compared to the Ising Hamiltonian. Furthermore, the mean approximation ratios of the QRAC Hamiltonian is higher than the Ising Hamiltonian with 12 nodes and 16 nodes. 

Fig.~\ref{fig:qrac_simulation}~(c) shows the mean approximation ratio with the candidate states obtained under depolarizing noise with error probability $1~\%$ on the controlled-NOT gate. By comparing Fig.~\ref{fig:qrac_simulation}~(a) and (c), we find that the mean approximation ratio of the QRAC Hamiltonian is robust to depolarizing noise as well. The simulation results under depolarizing noise captures the noise robustness of QRAO. We therefore explain the noise robustness of QRAO under the assumption of depolarizing noise.

\subsection{QRAO under depolarizing noise}
Depolarizing noise is where an $n$-qubit quantum state $\rho$ is mapped onto a linear combination of the unaffected state $\rho$ and the completely mixed state $I/2^n$. The noise model has a parameter $p\in [0, 1]$, which can be interpreted as the probability that the state $\rho$ remains unaffected by depolarizing noise. The state after a single application of depolarizing noise can be denoted as:
\begin{align}
\mathcal{D}_p(\rho) = p\rho + (1-p)\frac{I}{2^n}.
\label{eq:depo}
\end{align}
Depolarizing noise has the effect of ``shrinking'' the bloch sphere as shown in Fig.~\ref{fig:bloch}~(b). 
After $N$ applications of depolarizing noise, the resulting state becomes~\cite{grovertanaka}:
\begin{align}
\mathcal{D}^N_p(\rho) = p^N\rho + (1-p^N)\frac{I}{2^n}.
\label{eq:noisy_rho}
\end{align}

The robustness of the approximation ratio of Pauli rounding can be explained by the fact that the sign of the trace values are unaffected by depolarizing noise: 
\begin{align}
\Tr(P_j\mathcal{D}^N_p(\rho)) 
&= 
\Tr[P_j\left(p^N\rho + (1-p^N)\frac{I}{2^n}\right)]\nonumber \\
&= 
p^N\Tr[P_j\rho] + (1-p^N)\Tr[P_j\frac{I}{2^n}]\nonumber \\ 
&= p^N\Tr[P_j\rho],
\label{eq:pauli_abs_trace}
\end{align}
where $P_j$ is the Pauli matrix corresponding to the $j$-th node, $N$ the number of depolarizing noise applications, $n$ the number of qubits of $\rho$, and $\rho$ the candidate state of the Hamiltonian. While the sign remains unchanged, the absolute trace values decrease under depolarizing noise, causing the number of shots required to correctly estimate their sign to increase. Suppose that $|V|$ Pauli matrices are assigned to an $|V|$-node graph. To estimate the sign of the trace value corresponding to each node with error probability at most $\delta$, the minimum number of shots $S$ is derived in appendix A as
\begin{align*}
S\geq \frac{\ln(1/\delta)}{2\varepsilon^2},
\end{align*}
and the order of shots as
\begin{align*}
\mathcal{O}\left(\frac{\ln(|V|)}{\varepsilon^2}\right).
\end{align*}
Here, $\varepsilon > 0$ is defined by $\Pr(X_{ij} = 1) = 1/2 + \varepsilon$, where $X_{ij}$ denotes the measurement result of 0 or 1 corresponding to the $i$-th shot with respect to the Pauli matrix $P_j$ assigned to the $j$-th node. For $|V|$ Pauli matrices, the order of shots becomes
\begin{align}
\mathcal{O}\left(\frac{|V|\ln(|V|)}{\varepsilon^2}\right).
\label{eq:pauli_shots}
\end{align}
As seen in Eq.~(\ref{eq:pauli_shots}), the order of shots grows quadratically with the decrease of $\varepsilon$. Under depolarizing noise, $\varepsilon$ has the relation
\begin{align*}
\varepsilon = - p^N\Tr(P_j\rho)/2.
\end{align*} 
The minimum number of shots required under depolarizing noise can thus be written as 
\begin{align*}
S \geq 
\frac{4\ln(1/\delta)}{p^{2N}\Tr(P_j\rho)^2}.
\end{align*}
The number of shots required to correctly estimate all trace values correctly grows exponentially with the number of depolarizing noise applications $N$ and the number of nodes $|V|$, assuming that the QRAC Hamiltonian achieves the maximum compression rate where an $|V|$-node graph is encoded on $1/3$ qubits, and that the candidate state is obtained via an ansatz with $l$ linear entanglement layers under depolarizing noise. The ratio of the minimum number of shots required for the QRAC Hamiltonian to the Ising Hamiltonian is then given by $p^{\frac{3}{4}l|V|}$,
which indicates that estimating the correct configuration of all nodes with the same level of accuracy requires more shots with the Ising Hamiltonian than with the QRAC Hamiltonian.

The effect of depolarizing noise on the expected approximation ratio of Magic state rounding is as follows. Let $\rho_1$ and $\rho_3$ be density matrices corresponding to the Ising Hamiltonian $H_1$ and the QRAC Hamiltonian $H_3$ that satisfy $\Tr(H_1\rho_1) = \cut(m^*)$ and $\Tr(H_3\rho_3)\geq \cut(m^*)$, respectively. Without noise, the expected approximation ratio for $\rho_1$ is given by 
\begin{align}
\mathbb{E}(\gamma_1) = \frac{\Tr(H\rho_1)}{\cut(m^*)} = 1,
\end{align}
whereas the expected approximation ratio for $\rho_3$ via magic state rounding is given by
\begin{align}
\mathbb{E}(\gamma_3) 
=
\frac{
\mathbb{E}
\left[
\Tr\left\{
\mathcal{M}^{\otimes n_3}(\rho_3)H
\right\}
\right]
}{\cut(m*)}
\geq \frac{5}{9},
\end{align}
where $n_3$ represents the number of qubits of $\rho_3$. Now, let us consider parameterized circuits that output $\rho_1$ and $\rho_3$. We approximate the maximum-eigenvalue eigenstates by VQE under depolarizing noise by assuming that the parameterized circuits contain $N$ operations under the influence of depolarizing noise. The circuits that otherwise output $\rho_1$ and $\rho_3$ now output $\mathcal{D}^{N_1}_p(\rho_1)$ and $\mathcal{D}^{N_3}_p(\rho_3)$. The expected approximation ratio of magic state rounding corresponding to the resulting noisy states can be derived as
\begin{align}
\mathbb{E}(\gamma_1')
= p^{N_1} + (1-p^{N_1})\frac{|E|}{2\cut(m^*)}
\label{eq:qubo_bound}
\end{align}
for the Ising Hamiltonian and 
\begin{align}
\mathbb{E}(\gamma_3') &= 
\frac{
\mathbb{E}
\left[
\Tr\left\{
\mathcal{M}^{\otimes n_3}(\mathcal{D}^{N_3}_p(\rho_3))H
\right\}
\right]
}{\cut(m*)}\nonumber \\ 
&\geq \frac{5}{9}p^{N_3} + (1-p^{N_3})\frac{|E|}{2\cut(m^*)}
\label{eq:qrac_bound}
\end{align}
for the QRAC Hamiltonian. Note that the inequality $\mathbb{E}(\gamma_3') \geq \mathbb{E}(\gamma_1')$ holds only when $\cut(m^*)/|E| > 9/10$. 
\begin{figure}[H]
    \centering
    \hspace{-5mm}
    \includegraphics[width=88mm]{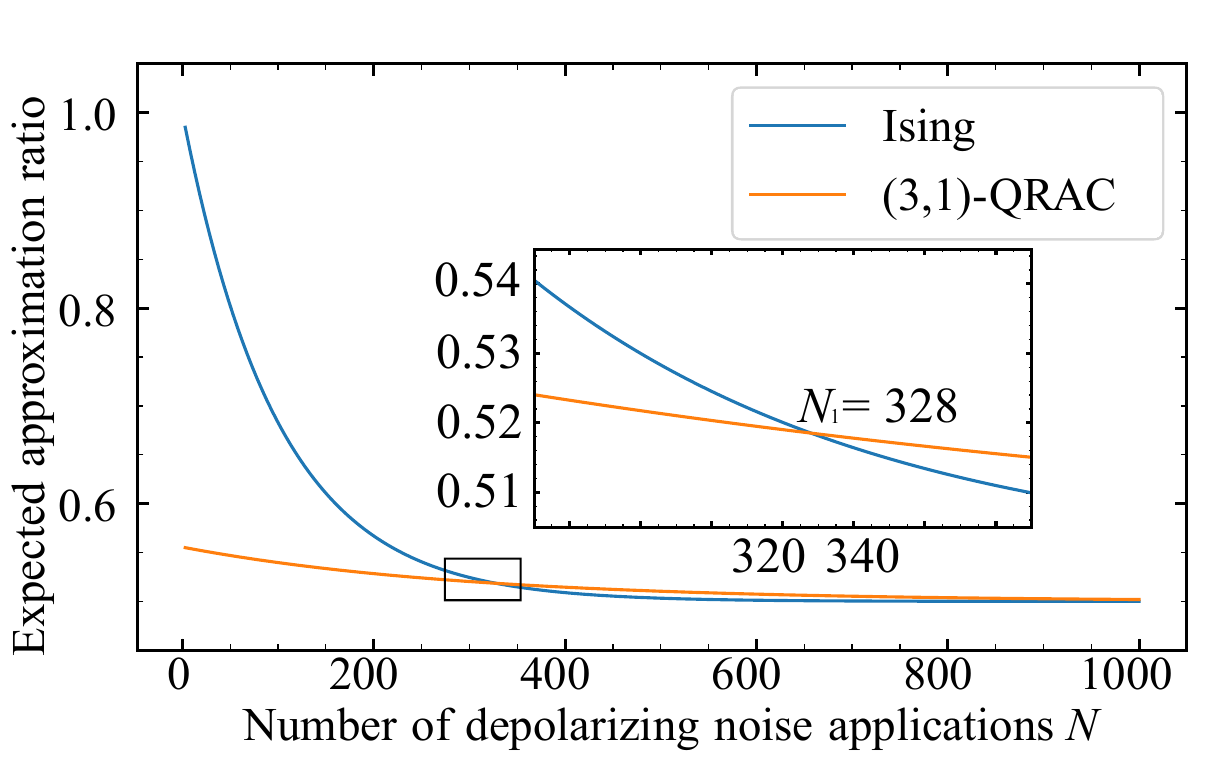}
    \caption{The expected approximation ratios of the Ising Hamiltonian and the QRAC Hamiltonian under depolarizing noise with success probability $p=0.99$ and the minimum number of depolarizing noise applications required to achieve $\frac{5}{9}p^{N_3} + (1-p^{N_3})/2\geq \mathbb{E}(\gamma_1')$}
    \label{fig:model_simulation}
\end{figure}
Considering the ideal case where the QRAC Hamiltonian requires 1/3 the number of qubits required with the Ising Hamiltonian, and that the graphs satisfy $|E| = \cut(m^*)$, we can simulate the expected approximation ratios of the Ising Hamiltonian and the QRAC Hamiltonian with respect to the number of noisy operations the candidate states have undergone. Fig.~\ref{fig:model_simulation}~(a) shows that there exists an $N_1$ where the lower bound of the expected approximation ratio of the QRAC Hamiltonian exceeds that of the Ising Hamiltonian under depolarizing noise with error probability $1~\%$. The assumptions here are that the graphs satisfy $|E|/\cut(m^*) > 9/10$ and that the QRAC candidate state $\rho_3$ goes thorough the depolarizing channel 1/3 the number of times the Ising Hamiltonian candidate state $\rho_1$ does. The intersection point $N_1$ decreases as the problem-size increases and the number of noise applications grows, or the success probability of each depolarizing noise application declines.

The robustness of the approximation ratio of the QRAC Hamiltonian compared to the Ising Hamiltonian can be explained by the fact that the QRAC Hamiltonian requires less qubits to encode the same problem than the Ising Hamiltonian. With less qubits, the amount of noisy operations in the PQC is reduced, leading to a candidate state that is less affected by noise. Additionally for Pauli rounding, the robustness is due to the sign of the trace value remaining unaffected under isotropic noise such as depolarizing noise.

\section{Conclusion}
In this paper, we have shown that the mean approximation of QRAO with the QRAC Hamiltonian is more robust to noise compared to the Ising Hamiltonian. We have observed that under fake noise, the mean approximation of the Ising Hamiltonian drops significantly under level of significance 5~\%, while the mean approximation ratio of the QRAC Hamiltonian does not. In addition, the mean approximation ratio of the QRAC Hamiltonian has been shown to exceed the mean approximation ratio of the Ising Hamiltonian as the problem-size increases, even in the absence of noise. We have shown that the mean approximation ratio of the QRAC Hamiltonian is higher in relation to its VQE energy ratio compared to the Ising Hamiltonian. Under the assumption of depolarizing noise, we have explained the robustness of the mean approximation ratio of QRAO by the fact that the sign of the trace values remain unaffected, and have explained the effect of depolarizing noise on the mean approximation ratio of QRAO via Magic state rounding by the fact that the candidate state of the QRAC Hamiltonian undergoes less noisy operations in VQE compared to the Ising Hamiltonian due to using less qubits. Finally, we have shown that the number of shots required to estimate the correct binary solution with the same level of accuracy is less for the QRAC Hamiltonian, and that the difference increases along with the problem-size and the level of noise. These facts indicate that the use of QRAO becomes an evermore realistic option as the problem-size increases under noise, because the mean approximation ratio of the QRAC Hamiltonian is expected to exceed the mean approximation ratio of the Ising Hamiltonian, and requires less qubits and shots.

\section*{Acknowledgment}
 N.Y. is
supported by MEXT Quantum Leap Flagship Program
(Nos. JPMXS0118067285, JPMXS0120319794). The views expressed are
those of the authors and do not reflect the official policy
or position of IBM or the IBM Quantum team.

\bibliographystyle{unsrt}
\bibliography{ref}

\begin{thebibliography}{10}

\bibitem{grotschel1995combinatorial}
Martin Grotschel and L{\'a}szl{\'o} Lov{\'a}sz.
\newblock Combinatorial optimization.
\newblock {\em Handbook of combinatorics}, 2(1541-1597):4, 1995.

\bibitem{azad2022solving}
Utkarsh Azad, Bikash~K Behera, Emad~A Ahmed, Prasanta~K Panigrahi, and Ahmed
  Farouk.
\newblock Solving vehicle routing problem using quantum approximate
  optimization algorithm.
\newblock {\em IEEE Transactions on Intelligent Transportation Systems}, 2022.

\bibitem{kurowski2023application}
Krzysztof Kurowski, Tomasz Pecyna, Mateusz Slysz, Rafa{\l} R{\'o}{\.z}ycki,
  Grzegorz Walig{\'o}ra, and Jan W\c{e}glarz.
\newblock Application of quantum approximate optimization algorithm to job shop
  scheduling problem.
\newblock {\em European Journal of Operational Research}, 310(2):518--528,
  2023.

\bibitem{Egger2020}
Daniel~J. Egger, Claudio Gambella, Jakub Marecek, Scott McFaddin, Martin
  Mevissen, Rudy Raymond, Andrea Simonetto, Stefan Woerner, and Elena Yndurain.
\newblock Quantum computing for finance: State-of-the-art and future prospects.
\newblock {\em IEEE Transactions on Quantum Engineering}, 1:1--24, 2020.

\bibitem{wang2020}
Samuel Marsh and Jingbo Wang.
\newblock Combinatorial optimization via highly efficient quantum walks.
\newblock {\em Phys. Rev. Research}, 2:023302, 2020.

\bibitem{vqe}
Alberto Peruzzo, Jarrod McClean, Peter Shadbolt, Man-Hong Yung, Xiao-Qi Zhou,
  Peter~J Love, Al{\'a}n Aspuru-Guzik, and Jeremy~L O’brien.
\newblock A variational eigenvalue solver on a photonic quantum processor.
\newblock {\em Nature communications}, 5(1):4213, 2014.

\bibitem{qaoa}
Edward Farhi, Jeffrey Goldstone, and Sam Gutmann.
\newblock A quantum approximate optimization algorithm.
\newblock {\em arXiv preprint arXiv:1411.4028}, 2014.

\bibitem{qaoa2}
G.~G. Guerreschi and A.~Y. Matsuura.
\newblock Qaoa for max-cut requires hundreds of qubits for quantum speed-up.
\newblock {\em Sci Rep}, 9:6903, 2019.

\bibitem{abbas2023quantum}
Amira Abbas, Andris Ambainis, Brandon Augustino, Andreas Bärtschi, Harry
  Buhrman, Carleton Coffrin, Giorgio Cortiana, Vedran Dunjko, Daniel~J. Egger,
  Bruce~G. Elmegreen, Nicola Franco, Filippo Fratini, Bryce Fuller, Julien
  Gacon, Constantin Gonciulea, Sander Gribling, Swati Gupta, Stuart Hadfield,
  Raoul Heese, Gerhard Kircher, Thomas Kleinert, Thorsten Koch, Georgios
  Korpas, Steve Lenk, Jakub Marecek, Vanio Markov, Guglielmo Mazzola, Stefano
  Mensa, Naeimeh Mohseni, Giacomo Nannicini, Corey O'Meara, Elena~Peña Tapia,
  Sebastian Pokutta, Manuel Proissl, Patrick Rebentrost, Emre Sahin, Benjamin
  C.~B. Symons, Sabine Tornow, Victor Valls, Stefan Woerner, Mira~L.
  Wolf-Bauwens, Jon Yard, Sheir Yarkoni, Dirk Zechiel, Sergiy Zhuk, and Christa
  Zoufal.
\newblock Quantum optimization: Potential, challenges, and the path forward,
  2023.

\bibitem{moll2018quantum}
Nikolaj Moll, Panagiotis Barkoutsos, Lev~S Bishop, Jerry~M Chow, Andrew Cross,
  Daniel~J Egger, Stefan Filipp, Andreas Fuhrer, Jay~M Gambetta, Marc Ganzhorn,
  et~al.
\newblock Quantum optimization using variational algorithms on near-term
  quantum devices.
\newblock {\em Quantum Science and Technology}, 3(3):030503, 2018.

\bibitem{tang2021cutqc}
Wei Tang, Teague Tomesh, Martin Suchara, Jeffrey Larson, and Margaret
  Martonosi.
\newblock Cutqc: using small quantum computers for large quantum circuit
  evaluations.
\newblock In {\em Proceedings of the 26th ACM International conference on
  architectural support for programming languages and operating systems}, pages
  473--486, 2021.

\bibitem{Tan2021qubitefficient}
Benjamin Tan, Marc-Antoine Lemonde, Supanut Thanasilp, Jirawat Tangpanitanon,
  and Dimitris~G. Angelakis.
\newblock Qubit-efficient encoding schemes for binary optimisation problems.
\newblock {\em {Quantum}}, 5:454, May 2021.

\bibitem{yano}
Hiroshi Yano, Yudai Suzuki, Kohei~M. Itoh, Rudy Raymond, and Naoki Yamamoto.
\newblock Efficient discrete feature encoding for variational quantum
  classifier.
\newblock {\em IEEE Transactions on Quantum Engineering}, 2:1--14, 2021.

\bibitem{qrao}
Bryce Fuller, Charles Hadfield, Jennifer~R Glick, Takashi Imamichi, Toshinari
  Itoko, Richard~J Thompson, Yang Jiao, Marna~M Kagele, Adriana~W
  Blom-Schieber, Rudy Raymond, et~al.
\newblock Approximate solutions of combinatorial problems via quantum
  relaxations.
\newblock {\em arXiv preprint arXiv:2111.03167}, 2021.

\bibitem{ambainis2002dense}
Andris Ambainis, Ashwin Nayak, Amnon Ta-Shma, and Umesh Vazirani.
\newblock Dense quantum coding and quantum finite automata.
\newblock {\em Journal of the ACM (JACM)}, 49(4):496--511, 2002.

\bibitem{qrao_entanglement}
Kosei Teramoto, Rudy Raymond, and Hiroshi Imai.
\newblock The role of entanglement in quantum-relaxation based optimization
  algorithms.
\newblock {\em arXiv preprint arXiv:2302.00429}, 2023.

\bibitem{qrao_tetrahedron}
Kosei Teramoto, Rudy Raymond, Eyuri Wakakuwa, and Hiroshi Imai.
\newblock Quantum-relaxation based optimization algorithms: Theoretical
  extensions.
\newblock {\em arXiv preprint arXiv:2302.09481}, 2023.

\bibitem{2024arXiv240302045K}
Ruho {Kondo}, Yuki {Sato}, Rudy {Raymond}, and Naoki {Yamamoto}.
\newblock {Recursive Quantum Relaxation for Combinatorial Optimization
  Problems}.
\newblock {\em arXiv e-prints}, page arXiv:2403.02045, March 2024.

\bibitem{holevo1973}
Holevo~A. S.
\newblock Bounds for the quantity of information transmitted by a quantum
  communication channel.
\newblock {\em Problemy Peredachi Informatsii}, 9:3--11, 1973.

\bibitem{hayashi20064}
Masahito Hayashi, Kazuo Iwama, Harumichi Nishimura, Rudy Raymond, and Shigeru
  Yamashita.
\newblock (4, 1)-quantum random access coding does not exist—one qubit is not
  enough to recover one of four bits.
\newblock {\em New Journal of Physics}, 8(8):129, 2006.

\bibitem{Liab_tr__2017}
O.~Liab{\o}tr{\o}.
\newblock Improved classical and quantum random access codes.
\newblock {\em Physical Review A}, 95(5), may 2017.

\bibitem{imamichi2018constructions}
Takashi Imamichi and Rudy Raymond.
\newblock Constructions of quantum random access codes.
\newblock In {\em Asian Quantum Information Symposium (AQIS)}, volume~66, 2018.

\bibitem{manvcinska2022geometry}
Laura Man{\v{c}}inska and Sigurd~AL Storgaard.
\newblock The geometry of {B}loch space in the context of quantum random access
  codes.
\newblock {\em Quantum Information Processing}, 21(4):1--16, 2022.

\bibitem{marques2022quantum}
Breno Marques and Rafael~A. da~Silva.
\newblock Quantum random access code in noisy channels, 2022.

\bibitem{qiskit}
{Qiskit contributors}.
\newblock Qiskit: An open-source framework for quantum computing, 2023.

\bibitem{Karp1972}
Richard~M. Karp.
\newblock {\em Reducibility among Combinatorial Problems}, pages 85--103.
\newblock Springer US, Boston, MA, 1972.

\bibitem{watanabe2023optimizing}
Hiroshi~C Watanabe, Rudy Raymond, Yu-ya Ohnishi, Eriko Kaminishi, and Michihiko
  Sugawara.
\newblock Optimizing parametrized quantum circuits with free-axis single-qubit
  gates.
\newblock {\em IEEE Transactions on Quantum Engineering}, 2023.

\bibitem{wada2022simulating}
Kaito Wada, Rudy Raymond, Yu-ya Ohnishi, Eriko Kaminishi, Michihiko Sugawara,
  Naoki Yamamoto, and Hiroshi~C Watanabe.
\newblock Simulating time evolution with fully optimized single-qubit gates on
  parametrized quantum circuits.
\newblock {\em Physical Review A}, 105(6):062421, 2022.

\bibitem{wada2022full}
Kaito Wada, Rudy Raymond, Yuki Sato, and Hiroshi~C Watanabe.
\newblock Full optimization of a single-qubit gate on the generalized
  sequential quantum optimizer.
\newblock {\em arXiv preprint arXiv:2209.08535}, 2022.

\bibitem{nft}
Ken~M Nakanishi, Keisuke Fujii, and Synge Todo.
\newblock Sequential minimal optimization for quantum-classical hybrid
  algorithms.
\newblock {\em Physical Review Research}, 2(4):043158, 2020.

\bibitem{ostaszewski2021structure}
Mateusz Ostaszewski, Edward Grant, and Marcello Benedetti.
\newblock Structure optimization for parameterized quantum circuits.
\newblock {\em Quantum}, 5:391, 2021.

\bibitem{grovertanaka}
Tomoki Tanaka, Yohichi Suzuki, Shumpei Uno, Rudy Raymond, Tamiya Onodera, and
  Naoki Yamamoto.
\newblock Amplitude estimation via maximum likelihood on noisy quantum
  computer.
\newblock {\em Quantum Information Processing}, 20(9):293, 2021.

\end{thebibliography}

\newpage
\appendix 
\section{Hoeffding-Chernoff bound for Pauli rounding}
Given a parameterized quantum circuit (PQC) with optimized parameters, apply the appropriate gate operation to all qubits and perform measurement in the computational basis to estimate the trace values corresponding to the Pauli matrix $X, Y$ or $Z$. The appropriate gate operation is the Hadamard gate $H$ for the Pauli matrix $X$, the Hadamard gate $H$ and then the phase gate $S$  for the Pauli matrix $Y$, and for the Pauli matrix $Z$, no gate operation is necessary.

Let us consider a graph with $|V|$ nodes, where the Pauli matrix corresponding to the $j$-th node is denoted $P_j$. A table of the measurement results across $S$ shots for the Pauli matrices $P_1\sim P_m$ may look like the following.
\begin{table}[H]
\centering
\caption{\small The $i$-th measurement result corresponding to the Pauli matrix $P_j$, where $j$ is the node number.}
\begin{tabular}{c|ccccc}
$i$ & $P_1$ & $\cdots$ & $P_j$ & $\cdots$ & $P_m$ \\ \hline \hline
1 & 1 & $\cdots$ & 1 & $\cdots$ & 0 \\
2 & 0 & $\cdots$ & 1 & $\cdots$ & 1 \\
$\vdots$ & $\vdots$ & $\vdots$ & $\vdots$ & $\vdots$ & $\vdots$\\ 
$i$ & $X_{i1}$ & $\cdots$ & $X_{ij}$ & $\cdots$ & $X_{im}$ \\
$\vdots$ & $\vdots$ & $\vdots$ & $\vdots$ & $\vdots$ & $\vdots$\\ 
$n$ & 0 & $\cdots$ & 0 & $\cdots$ & 1 \\ \hline 
$h_1^j=\sum_{i=1}^{S}X_{ij}$ & $h_1^1$ & {} & $h_1^j$ & {} & $h_1^{|V|}$\\ \hline
\end{tabular}
\end{table}
\noindent After $S$ shots, we have $S$ measurement results for each node $j$ as $X_{1j}, X_{2j}, \ldots, X_{Sj}$. The trace value is estimated as
\begin{align*}
\Tr(P_j\rho)
&= 
2
\left(
1 - 
\frac{\sum_{i=1}^{S}X_{ij}}{S}
\right)
- 1\\
&= 
1 - 2\frac{\sum_{i=1}^{S}X_{ij}}{S}\\
&=
1 - 2\frac{h_1^j}{S}.
\end{align*}
Assume that $\Pr[X_{i, j} = 1] = \frac{1}{2} + \varepsilon$, where $\Pr[X_{i,j} = 1]$ is the expectation value of the measurement outcome of $X_{i,j}$, and $\varepsilon > 0$. Note that under this assumption, the trace value is always negative. We estimate the sign of the trace value incorrectly when $h_1^j = \sum_{i=1}^{S}X_{ij}\leq S/2$, which leads to a positive sign. 
\paragraph{Chernoff-Hoeffding bound} Let $X_{1j}, X_{2j}, \ldots, X_{Sj}$ be independent random variables in $\{0, 1\}$ with $\Pr[X_{i,j} = 1] = p_j$. Let $h_1^j = \sum_{i=1}^{S}X_{ij}$, $\mu_j = E[h_1^j] = Sp_j$. Then, for any $\lambda > 0$ it holds:
\begin{align*}
\Pr[
h_1^j\leq \mu_j - \lambda
]&\leq \exp(-\frac{2\lambda^2}{n}).
\end{align*}
By substituting $\mu$ with $n(1/2 + \varepsilon)$ and $\lambda$ with $\varepsilon S$, we obtain 
\begin{align*}
\Pr[
h_1^j\leq \mu_j - \lambda
]
&= 
\Pr[
h_1^j\leq S\left(\frac{1}{2} + \varepsilon\right) - \varepsilon S
]\\
&=
\Pr[
h_1^j\leq \frac{S}{2}
] \leq \exp(-2S\varepsilon^2).
\end{align*}
Now, we would like to bound the probability of estimating the sign of the trace value incorrectly with $\delta$ as
\begin{align*}
\Pr[
h_1^j\leq \frac{S}{2}
] \leq \exp(-2S\varepsilon^2)
\leq \delta.
\end{align*}
From the above inequality, it follows that 
\begin{align*}
S\geq \frac{\ln(1/\delta)}{2\varepsilon^2}.
\end{align*}
The probability of correctly estimating the sign of the trace values corresponding to all Pauli matrices is given by $(1 - \delta)^{|V|}$, which should be close to 1.
From Bernoulli's inequality, we have
\begin{align*}
\alpha \leq (1 - \delta)^{|V|} \leq \exp(-\delta |V|),
\end{align*}
from which can be derived $\delta\leq \ln(\alpha)/|V|$. Now, we have 
\begin{align*}
S = \mathcal{O}\left(\frac{\ln(|V|)}{\varepsilon^2}\right).
\end{align*}
For $m$ Pauli matrices, the number of shots becomes 
\begin{align*}
\mathcal{O}\left(\frac{|V|\ln(|V|)}{\varepsilon^2}\right).
\end{align*}

In the case where the candidate state $\rho$ is affected by depolarizing noise, the trace value becomes $\Tr(P_j\mathcal{D}_p^N(\rho)) = p^N\Tr(P_j\rho)$. With this in mind, the value of $\varepsilon$ can be derived as $\varepsilon = - p^N\Tr(P_j\rho)/2$. Note that the trace value is assumed to be negative.

\end{document}